\documentclass[10pt, twocolumn, twoside]{IEEEtran}
\ifCLASSINFOpdf
\else
\fi

\usepackage{cite}
\usepackage{graphics} 
\usepackage{epsfig} 
\usepackage[cmex10]{amsmath}
\usepackage{amssymb}  
\usepackage{citesort}
\usepackage{amsfonts}
\usepackage{amsthm}

\usepackage{booktabs}
\usepackage{wasysym}
\allowdisplaybreaks[1]
\newcommand{\ra}[1]{\renewcommand{\arraystretch}{#1}}

\newtheorem{proposition}{Proposition}
\makeatletter
\begin{document}
%
\title{A Blind Time-Reversal Detector in the Presence of Channel Correlation}
%
%
%

\author{Zhong~Zheng,~\IEEEmembership{Student~Member,~IEEE,}
        Lu~Wei,~\IEEEmembership{Student~Member,~IEEE,}
        Jyri~H\"{a}m\"{a}l\"{a}inen,~\IEEEmembership{Member,~IEEE,}
        and~Olav~Tirkkonen,~\IEEEmembership{Member,~IEEE}
\thanks{This work was supported by the Academy of Finland and the Finnish Funding Agency for Technology and Innovation (TEKES).

The authors are with the Department of Communications and Networking, Aalto University, 00076 Aalto, Finland (e-mail: \{zhong.zheng, lu.wei, jyri.hamalainen, olav.tirkkonen\}@aalto.fi). J. H\"am\"al\"ainen is also with Ericsson Oulu R\&D Center, Finland.}}

\maketitle

\begin{abstract}
A blind target detector using the time reversal transmission is proposed in the presence of channel correlation. We calculate the exact moments of the test statistics involved. The derived moments are used to construct an accurate approximative Likelihood Ratio Test~(LRT) based on multivariate Edgeworth expansion. Performance gain over an existing detector is observed in scenarios with channel correlation and relatively strong target signal.
\end{abstract}


%
\IEEEpeerreviewmaketitle

\section{Introduction}
%
%
%
%
\IEEEPARstart{T}{ime} Reversal~(TR) is a waveform transmission method that focuses the transmitted energy in dispersive \mbox{medium -- the channel}~\cite{fink_1992}. It utilizes channel reciprocity and obtains the channel state information by sending a probing signal. The backscattered signal is then time-reversed and retransmitted. The TR signal is shown to be optimal in the sense that the transmission realizes a matched filter to the propagation transfer function~\cite{fink_1992}. The concept of~TR was originally developed for optical and acoustic applications, and it is recently introduced as a detection technique in the electromagnetic domain~\cite{jin_2009,jin_2010,dono_2012}, where the target to be detected is embedded in stationary random multipath scatterers.

In~\cite{jin_2009,jin_2010}, the authors assumed that the multipath channel or the channel response signal can be ideally estimated using probing snapshots. However, the assumption of a perfectly known channel or a noise-free signal may not be realistic for practical systems due to e.g. measurement noise. Estimation accuracy depends on the number of snapshots, which is limited by the coherence time/frequency of the channel, and the sampling rate of the system~\cite{jin_2010}. To avoid channel estimation, the authors in~\cite{dono_2012} considered a blind~TR detector, which utilizes only the distribution of the multipath channels. The likelihood ratio test for the TR detector was derived assuming statistical independence between the two consecutive transmissions. However, this assumption is not valid if the transmissions are within the coherence time of the multipath channel. Using existing detectors in such a scenario will induce performance loss.

Despite the practical needs to understand TR detection in the presence of channel correlation, the results in this direction are scarce. To address this challenge, we propose a blind TR detector that admits a general correlation structure between the TR channels. A closed-form approximation to the corresponding likelihood ratio is proposed using the multivariate Edgeworth expansion. The approximation is constructed via the derived exact moments of the underlying statistics. Numerical simulations show that the proposed detector outperforms the detector in~\cite{dono_2012} by exploiting the TR channel correlations.

\section{Blind Time Reversal Detection}\label{sec_tr}

We consider blind detection of a point target in the presence of multipath scatterers as studied in~\cite{dono_2012}. The detection system sends $Q$ probing signals in the spectral domain at the frequencies $\omega_q$, $q\in[1,\ Q]$. The sampling frequencies are chosen such that each frequency bin is separated by the coherence bandwidth of the channel and the spectral samples are statistically independent. The multipath channel at $\omega_q$ induced by the scatterers is modeled by a wide sense stationary process. We denote the channels experienced by the probing signal and the retransmission as $C_p(\omega_q)$ and $C_r(\omega_q)$, respectively. The channel response of the point target is captured by a deterministic response~$T$ and the probing signal at $\omega_q$ is denoted as $S(\omega_q)$. Note that here we consider a general correlation structure between $C_p(\omega_q)$ and $C_r(\omega_q)$ instead of statistical independence assumed in~\cite{dono_2012}. As a result, knowledge of channel coherence time is no longer required. In such a scenario, the detector of~\cite{dono_2012} suffers performance loss as will be shown in Section~\ref{sec_result}.

After transmitting the probing signal~$S(\omega_q)$, we write the frequency response as
\begin{equation*}\label{eq_z}
Z(\omega_q)=(T+C_p(\omega_q))S(\omega_q)+V_p(\omega_q),
\end{equation*}
where $V_p(\omega_q)$ is the measurement noise which is distributed as a zero-mean complex Gaussian random variable with Power Spectral Density~(PSD)~$\sigma_v^2$. Hereafter, we denote~$V_p(\omega_q)\sim\mathcal{CN}(0,\sigma_v^2)$~\cite{pinin_1996}. In this paper, we use a white probing signal such that $S(\omega_q)=\sqrt{E_s/Q}$
with a transmit power $E_s$. The received signal~$Z(\omega_q)$ is then time-reversed or, equivalently, phase-conjugated in the frequency domain and scaled to obtain the TR signal,~\mbox{$S_{TR}(\omega_q)=k Z(\omega_q)^*$}, where $k=\sqrt{E_s/\sum_{q=1}^Q|Z(\omega_q)|^2}$ is a energy normalization factor.
The value of $k$ is shown to be approximately deterministic with a relatively small variance compared to its expected value~\cite{jin_2010}. The TR signal $S_{TR}(\omega_q)$ is subsequently transmitted and the channel response of the retransmission is calculated by
\begin{align}
Z_{TR}(\omega_q)&=(T+C_r(\omega_q))S_{TR}(\omega_q)+V_r(\omega_q)\nonumber\\
&=X(\omega_q)Y(\omega_q)^*+V_r(\omega_q)\label{eq_ztr},
\end{align}
where $X(\omega_q)=T+C_r(\omega_q)$, $Y(\omega_q)=S_{TR}(\omega_q)^*$ and $V_r(\omega_q)\sim\mathcal{CN}(0,\sigma_v^2)$ is the measurement noise of the retransmission. In blind TR detection, the channels $C_p(\omega_q)$ and $C_r(\omega_q)$ will not be estimated by the detector and are only known by their statistical distributions. Therefore, a hypothesis test can be formulated as follows: in the null hypothesis~$\mathcal{H}_0$, the target is not present and $T=0$; in the alternative hypothesis~$\mathcal{H}_1$, $|T|>0$.

We assume the channels $C_p(\omega_q)$ and $C_r(\omega_q)$ admit a bivariate zero-mean complex Gaussian distribution with a common PSD~$P_c(\omega_q)$. The correlation coefficient~$\rho_c$ between $C_p(\omega_q)$ and $C_r(\omega_q)$ is defined as~$\rho_c=\mathbb{E}[C_p(\omega_q) C_r(\omega_q)^*]/P_c(\omega_q)$,
where the notation $(\cdot)^*$ is the complex conjugate. In practical systems, the channel statistics can be estimated by taking snapshots of channel samples and replacing the statistical expectation by the sample mean. The measurement noise~$V_p(\omega_q)$ and $V_r(\omega_q)$ are independent of each other and the multipath channels. If we ignore the noise term~$V_r(\omega_q)$ in~(\ref{eq_ztr}), $Z_{TR}(\omega_q)$ is distributed as the product of two complex Gausian random variables with 
\begin{align}
X(\omega_q)&\sim\mathcal{CN}(T,P_c(\omega_q))\label{eq_rvx},\\ Y(\omega_q)&\sim\mathcal{CN}\left(kT\sqrt{\frac{E_s}{Q}}, k^2\left(P_c(\omega_q)\frac{E_s}{Q}+\sigma_v^2\right)\right)\label{eq_rvy}.
\end{align}
By definition, the random variables~$X(\omega_q)$ and~$Y(\omega_q)$ are jointly complex Gaussian distributed with a correlation coefficient calculated as
\begin{equation}\label{eq_rho}
\rho=\frac{\rho_c^*}{\sqrt{1+\sigma_v^2 Q/(P_c(\omega_q)E_s)}}.
\end{equation}

To clarify the considered problem, we introduce the random variable~$\mathcal{P}(\omega_q)=X(\omega_q)Y(\omega_q)^*$ and denote its corresponding PDF in the complex plane as~$f_{\mathcal{P}_q}(p_1, p_2; T)$. The LRT of the blind TR detection is calculated by
\begin{equation}\label{eq_lr}
l=\prod_{q=1}^Q \frac{f_{\mathcal{P}_q}(p_1,p_2;T)}{f_{\mathcal{P}_q}(p_1,p_2;0)}\overset{\mathcal{H}_{1}}{\underset{\mathcal{H}_{0}}{\gtrless}}l_0,
\end{equation}
$l_0$ being a threshold. In the next section, we first derive the characteristic function of the product~$\mathcal{P}(\omega_q)$. Based on this, an exact expression is obtained for $f_{\mathcal{P}_q}(p_1,p_2;0)$ and an asymptotic approximation is given for $f_{\mathcal{P}_q}(p_1,p_2;T)$.

\section{Correlated Time Reversal Channel}\label{sec_stat}

\subsection{Characteristic Function}
We first derive the characteristic function~$\psi_{\mathcal{P}}(t)$~($t\in\mathbb{C}$) of the product~$\mathcal{P}(\omega_q)=X(\omega_q)Y(\omega_q)^*$. For notational simplicity, the frequency variable~$\omega_q$ is hereafter dropped. Recalling equations~(\ref{eq_rvx})-(\ref{eq_rho}), the joint PDF of $X$ and $Y$ is given by~\cite{pinin_1996} as~$f_{X,Y}(x,y)=\frac{\exp\left\{ -g(x,y)/(1-|\rho|^2) \right\} }{ \pi^2 \sigma_X^2 \sigma_Y^2 (1-|\rho|^2) }$
where $g(x,y)=\left|\frac{x-\mu_X}{\sigma_X}\right|^2
						 +\left|\frac{y-\mu_Y}{\sigma_Y}\right|^2 -2\Re[\rho^*\frac{x-\mu_X}{\sigma_X}\frac{y^*-\mu_Y^*}{\sigma_Y}]$ and $\mu_i$, $\sigma_i$ ($i\in\{X,Y\}$) refer to the mean and variance of the corresponding random variable. Given~$Y$, $X$ is conditionally complex Gaussian distributed with mean $\mu_{X|Y}=\mu_X+\rho(y-\mu_Y)\sigma_X/\sigma_Y$ and variance $\sigma_{X|Y}^2=\sigma_X^2(1-|\rho|^2)$. Denote the real and imaginary parts of~$\mathcal{P}$ by~$\mathcal{P}_1$ and $\mathcal{P}_2$. It is straightforward to show that~$\mathcal{P}_1$ and~$\mathcal{P}_2$ conditioned on~$Y$ are conditionally independent. They follow conditional Gaussian distributions
\begin{align*}
\mathcal{P}_1|Y~\sim\mathcal{N}\left(\Re\left[y^*\mu_{X|Y}\right], \sigma_{X|Y}^2|y|^2/2\right),\\
\mathcal{P}_2|Y~\sim\mathcal{N}\left(\Im\left[y^*\mu_{X|Y}\right], \sigma_{X|Y}^2|y|^2/2\right).
\end{align*}
Therefore, the conditional characteristic function of $\mathcal{P}$ given $Y$ is expressed as~\cite{pinin_1996}
\begin{align}\label{eq_ccf}
\psi_{\mathcal{P}|Y}(t|y)&=\mathbb{E}\left[\exp\left\{i \Re\left[t^*\mathcal{P}\right]\right\}|Y=y\right]\nonumber\\
&=\exp\Bigl\{i \Re[t^*y^*\mu_{X|Y}]-\frac{1}{4}\sigma_{X|Y}^2|y|^2|t|^2\Bigr\}.
\end{align}
The marginal PDF of $Y$ is given by
\begin{equation}\label{eq_pdfy}
f_{Y}(y)=1/(\pi\sigma_{Y}^2)\exp\left\{-|y-\mu_Y|^2/\sigma_Y^2\right\},\quad y\in\mathbb{C}.
\end{equation}
The characteristic function of $\mathcal{P}$ can be now obtained by direct integration of~(\ref{eq_ccf}) over the marginal PDF (\ref{eq_pdfy}) as
\begin{align}\label{eq_cf1}
\psi_{\mathcal{P}}(t)=\int_{y\in\mathbb{C}}\psi_{\mathcal{P}|Y}(t|y)f_{Y}(y)\,\mathrm{d}y.
\end{align}
Substituting~(\ref{eq_ccf}) and~(\ref{eq_pdfy}) into~(\ref{eq_cf1}), we obtain
\begin{align}
&\psi_{\mathcal{P}}(t)=\exp\{-|\mu_Y|^2/\sigma_Y^2\}/(\pi\sigma_Y^2)\nonumber\\
&\times\int_{y\in\mathbb{C}}\exp\left\{-\left(\frac{1}{\sigma_Y^2}+\frac{\sigma_X^2|t|^2}{4(1-|\rho|^2)^{-1}}-i\Re[t^*\rho]\frac{\sigma_X}{\sigma_Y}\!\right)|y|^2\right.\nonumber\\
&\left.+\frac{2\Re[\mu_Y y^*]}{\sigma_Y^2}+i\Re\left[t^*\left(\mu_X-\rho\frac{\sigma_X}{\sigma_Y}\mu_Y\right)y^*\right]\right\}\,\mathrm{d}y.\label{eq_cf2}
\end{align}
Applying~\cite[eq. (3.323/2)]{gradshteyn_2007} and integrating (\ref{eq_cf2}) over real and imaginary parts of $y$, we get
\begin{align}\label{eq_cf3}
\psi_{\mathcal{P}}(t)&=\frac{1}{G(t)}\exp\left\{-\frac{|\mu_X|^2\sigma_Y^2+|\mu_Y|^2\sigma_X^2}{4 G(t)}|t|^2\right.\nonumber\\
&\left.+\frac{\sigma_X\sigma_Y\Re[\mu_X^*\mu_Y\rho]}{2G(t)}|t|^2+\frac{i\Re[\mu_X^*\mu_Y t]}{G(t)}\right\},
\end{align}
where $G(t)=1+\frac{1}{4}\sigma_X^2\sigma_Y^2(1-|\rho|^2)|t|^2-i\sigma_X\sigma_Y\Re[t^*\rho]$.

\subsection{Joint PDF}\label{sec_stat_jpdf}
Based on~(\ref{eq_cf3}), we now calculate the joint PDF~$f_{\mathcal{P}}(p_1,p_2;0)$ under the null hypothesis $\mathcal{H}_0$. When $T=0$, $\mu_X=\mu_Y=0$ and~(\ref{eq_cf3}) is reduced to $\psi_{\mathcal{P}}(t)=1/G(t)$. Applying the inverse transform of characteristic function, the joint PDF becomes
\begin{align}\label{eq_f0}
f_{\mathcal{P}}&(p_1,p_2;0)=\frac{1}{(2\pi)^2}\int_{t\in\mathbb{C}}\frac{\exp\{-i \Re[t^*p]\}}{G(t)}\,\mathrm{d}t\nonumber\\
&=\frac{2}{\pi\sigma_X\sigma_Y c}\exp\left\{\frac{2\Re[\rho^*p]}{c}\right\} K_0\left(\frac{2|p|}{c}\right),
\end{align}
where $p=p_1+i p_2$ and $c=\sigma_X\sigma_Y(1-|\rho|^2)$. Here, the function~$K_0(\cdot)$ is the modified Bessel function of the second kind~\cite[eq. (8.432/6)]{gradshteyn_2007}. The second equality of~(\ref{eq_f0}) is obtained by using~\cite[eq. (3.354/5)]{gradshteyn_2007} and the definition of~$K_0(\cdot)$.

Next, we derive an asymptotic approximation to the joint PDF~$f_{\mathcal{P}}(p_1,p_2;T)$ using the multivariate Edgeworth expansion. The Edgeworth expansion was considered as an extension to the central limit theorem and developed in the form of a moment series expansion weighted by the Gaussian PDFs~\cite{bhatt_1976}. This method is especially useful when the random variable of interest is approximately Gaussian and its moments are easy to obtain. For the problem at hand, we can prove that~$f_{\mathcal{P}}(p_1,p_2;T)$ converges to a Gaussian PDF as $\mu_X$ and $\mu_Y$ go to infinity. Motivated by this fact, we give the Edgeworth approximation to $f_\mathcal{P}(p_1,p_2;T)$ based on the closed-form expressions for the joint moments of $\mathcal{P}_1$ and $\mathcal{P}_2$. 

Let $W=(\mathcal{P}-\mu_\mathcal{P})/\sigma_\mathcal{P}$, where~$\mu_\mathcal{P}=\mu_X\mu_Y^*+\rho\sigma_X\sigma_Y$ and~$\sigma_\mathcal{P}^2=|\mu_X|^2\sigma_Y^2+|\mu_Y|^2\sigma_X^2+\sigma_X^2\sigma_Y^2$. The characteristic function of $W$ reads
\begin{equation}\label{eq_w}
\psi_W(t)=\exp\left\{-i\Re\left[\mu_\mathcal{P}t^*\right]/\sigma_\mathcal{P}\right\}\psi_{\mathcal{P}}\left(t/\sigma_\mathcal{P}\right)\!.
\end{equation}
We denote $\delta_X=\mu_X/\sigma_X$ and $\delta_Y=\mu_Y/\sigma_Y$ and notice that as~$|\delta_X|$ and $|\delta_Y|$ go to infinity (\ref{eq_w}) reduces to~$\exp\{-|t|^2/4\}$, which is the characteristic function of the standard complex Gaussian random variable. Thus, the considered variate $\mathcal{P}$ is approximately complex Gaussian with mean $\mu_\mathcal{P}$ and variance $\sigma_\mathcal{P}^2$.

Before constructing the multivariate Edgeworth expansion for $f_\mathcal{P}(p_1,p_2;T)$, we need the following two propositions:
\begin{proposition}\label{prop_1}
Let $X=\mu_X+V_X$ and $Y=\mu_Y+V_Y$, where $V_X\sim\mathcal{CN}(0,\sigma_X^2)$ and $V_Y\sim\mathcal{CN}(0,\sigma_Y^2)$ with the correlation $\mathbb{E}[V_X V_Y^*]=\rho\sigma_X\sigma_Y$. Denote~$A_{t,i}=\{\underbrace{V_X,\cdots, V_X}_{i}, \underbrace{V_Y,\cdots, V_Y}_{t-i}\}$ and $B_{t,j}=\{\underbrace{V_X^*,\cdots, V_X^*}_{t-j}, \underbrace{V_Y^*,\cdots, V_Y^*}_j\}$.
The joint moment $\mathcal{M}_{m,n}=\mathbb{E}[\mathcal{P}^m(\mathcal{P}^n)^*]$ is given by
\begin{align*}\label{eq_jcm}
&\hspace{-1.5ex}\mathcal{M}_{m,n}=\sum_{t=0}^{m+n}\sum_{i=0}^t\sum_{j=0}^t{m\choose i}{m\choose j}{n\choose t-j}{n\choose t-i}\nonumber\\
&\times\! \mu_X^{m-i}(\mu_Y^{m-j})^*(\mu_X^{n-t+j})^*\mu_Y^{n-t+i}\mathop{\sum}_{\pi\in\Omega}\prod_{k=1}^t\mathbb{E}[A_{t,i}^{(\pi^{(k)})}B_{t,j}^{(k)}],
\end{align*}
where $(\cdot)^{(i)}$ denotes the $i$-th element of the corresponding vector, $\pi$ defines a permutation of the integers~$1,\ldots,t$ and~$\Omega$ is the set of all~$t!$ distinct permutations.
\end{proposition}
The proof of Proposition~\ref{prop_1} is a direct application of the moment theorem in~\cite{reed_1962}, which is omitted here.


\begin{proposition}\label{prop_2}
The joint moments of $\mathcal{P}_1$ and $\mathcal{P}_2$ are
\begin{equation*}
\left(
\begin{array}{c}
\mathbb{E}[\mathcal{P}_1^{m}\mathcal{P}_2^{0}]\\
\mathbb{E}[\mathcal{P}_1^{m-1}\mathcal{P}_2^{1}]\\
\vdots\\
\mathbb{E}[\mathcal{P}_1^{0}\mathcal{P}_2^{m}]
\end{array}
\right)=\mathbf{J}_m^{-1}\left(
\begin{array}{c}
\mathcal{M}_{m,0}\\
\mathcal{M}_{m-1,1}\\
\vdots\\
\mathcal{M}_{0,m}
\end{array}
\right),
\end{equation*}
where $\mathbf{J}_m^{(k+1,\ l+1)}$, the element of the $(k+1)$-th row and $(l+1)$-th column of matrix $\mathbf{J}_m$, is given by
\begin{equation}\label{eq_j1}
\mathbf{J}_m^{(k+1,\ l+1)}=\sum_{h=0}^{\lfloor l/2\rfloor}i^{l-2h}{m-2k\choose l-2h}{k\choose h}, 
\end{equation}
where $\lfloor a\rfloor$ denotes the largest integer less than $a$.
\end{proposition}
\begin{IEEEproof}
First consider the case $m\ge 2k$, 
\begin{align*}\label{eq_j2}
&\mathcal{M}_{m-k,k}=\mathbb{E}\left[\left(\mathcal{P}_1+i\mathcal{P}_2\right)^{m-2k}\left(\mathcal{P}_1^2+\mathcal{P}_2^2\right)^k\right]\\
&=\sum_{j=0}^{m-2k}\sum_{h=0}^{k}i^j{m-2k\choose j}{k\choose h}\mathbb{E}\left[\mathcal{P}_1^{m-j-2h}\mathcal{P}_2^{j+2h}\right].
\end{align*}
Summing over $l$ with $l=j+2h$, the coefficient of the~$l$-th term achieves~(\ref{eq_j1}). In the same manner, (\ref{eq_j1}) can be also proved when $m<2k$.
\end{IEEEproof}

Following the procedures in~\cite{bhatt_1976}, the Edgeworth expansion~$f_s(p_1,p_2)$ of the joint PDF~$f_\mathcal{P}(p_1,p_2;T)$ can be represented as
\begin{equation}\label{eq_f1}
f_\mathcal{P}(p_1,p_2;T)\approx\phi(p_1,p_2)+\sum_{j=1}^{s-2}L_j\left(-\phi;\chi_\nu\right)(p_1,p_2),
\end{equation}
where $\phi(p_1,p_2)$ is the bivariate Gaussian PDF with mean $\mathbf{\mu}=[\mu_{\mathcal{P}_1}, \mu_{\mathcal{P}_2}]^T$ and covariance matrix~$\mathbf{R}$ of $\mathcal{P}_1$ and $\mathcal{P}_2$. Here $\chi_\nu$ with $\nu=[\nu_1,\nu_2]$ refers to the joint cumulant of $\mathcal{P}_1$ and $\mathcal{P}_2$. These can be readily calculated by the mapping between joint cumulants and moments, which are computed by Proposition~\ref{prop_1} and~\ref{prop_2}. It is noted that the expansion~(\ref{eq_f1}) uses up-to $s$-th order joint cumulants of $\mathcal{P}_1$ and $\mathcal{P}_2$ and matches the first~$(s-2)$ moments between $\mathcal{P}$ and the approximation~(\ref{eq_f1}). The term $L_j\left(-\phi;\chi_\nu\right)(p_1,p_2)$ defines the Cram$\acute{\mathrm{e}}$r Edgeworth polynomial $\widetilde{L_j}(t_1,t_2;\chi_\nu)$ with each $t_1^{\nu_1}t_2^{\nu_2}$ replaced by $H(p_1,p_2;\nu,\mathbf{R}^{-1})\phi(p_1,p_2)$, where $H$ denotes the multivariate Hermite polynomials~\cite{skov_1986}. The Cram$\acute{\mathrm{e}}$r Edgeworth polynomials $\widetilde{L_j}(t_1,t_2;\chi_\nu)$ can be generated by the formal identity between two power series~\cite{bhatt_1976}
\begin{align*}
\sum_{r=1}^\infty \widetilde{L_j}(t_1,t_2;\chi_\nu)u^r=\sum_{m=1}^\infty\frac{1}{m!}\left(\sum_{r=1}^\infty\frac{\chi_{r+2}(t_1,t_2)}{(r+2)!}u^r\right)^m,
\end{align*}
where $\chi_s(t_1,t_2)=\sum_{\nu_1+\nu_2=s}\frac{s!}{\nu_1!\nu_2!}\chi_\nu t_1^{\nu_1}t_2^{\nu_2}$.

\section{Numerical Results}\label{sec_result}
In this section we investigate the accuracy of the derived approximation to $f_\mathcal{P}(p_1,p_2; T)$ by simulations. Then we compare the performance of the proposed detector with the one studied in~\cite{dono_2012}.

\subsection{Approximation to $f_\mathcal{P}(p_1,p_2; T)$}

\begin{figure}[!t]
\centering
\includegraphics[width=2.4in]{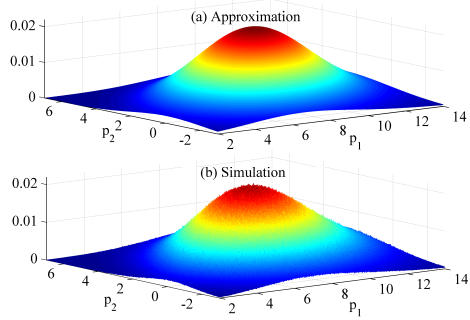}
\caption{Plot of two-dimensional PDF $f_\mathcal{P}(p_1,p_2; T)$ for $\mu_X=2+2.5i$, $\mu_Y=2.1+1.8i$, $\sigma_X=\sigma_Y=1$ and $\rho=0.3+0.3i$. (a) approximation~(\ref{eq_f1}) with $s=6$ and (b) simulation. The mean square error of approximation is $9.96\times10^{-7}$.} \label{fig_cpdf}
\end{figure}

In Fig.~\ref{fig_cpdf} we plot the approximative PDF~(\ref{eq_f1}) and compare with the simulated~$f_\mathcal{P}(p_1,p_2; T)$ assuming $\mu_X=2+2.5i$, $\mu_Y=2.1+1.8i$, $\sigma_X=\sigma_Y=1$ and $\rho=0.3+0.3i$. Fig.~\ref{fig_cpdf} shows that the Edgeworth approximation achieves a good agreement with the simulation using up-to $6$-th order joint cumulants of $\mathcal{P}_1$ and $\mathcal{P}_2$. The approximation is less accurate near the origin since the function $f_\mathcal{P}(p_1,p_2; T)$ has a simple singularity at the point~$(0,0)$. 

To further illustrate the accuracy of the approximation, in Table~\ref{tab_mse} we tabulate the Mean Square Error~(MSE) of the approximative PDF~(\ref{eq_f1}), defined as $\int_{\{p_1,p_2\}\in\mathbb{R}^2}|f_s(p_1,p_2)-\widetilde{f_\mathcal{P}}(p_1,p_2;T)|^2\,\mathrm{d}\widetilde{F_\mathcal{P}}(p_1,p_2;T)$, where $\widetilde{f_\mathcal{P}}$ and $\widetilde{F_\mathcal{P}}$ are empirical PDF and CDF of $\mathcal{P}$. In addition to the MSE calculated with the parameters used for Fig.~\ref{fig_cpdf}, we also calculate the MSEs with both $\mu_X$ and $\mu_Y$ having $3$~dB decrease and increase, respectively. The results show that the accuracy of the approximation~(\ref{eq_f1}) is improved as the magnitudes of $\mu_X$ and $\mu_Y$ increase, which is in line with the analysis in Section~\ref{sec_stat_jpdf}. This conclusion has been also verified with other parameter combinations.

\begin{table}[t!]
\caption{Mean square error of approximative distribution~(\ref{eq_f1})}\centering
\ra{1}
\begin{tabular}{@{}llll@{}}
\toprule
$\begin{array}{c}
\mu_X\\
\mu_Y
\end{array}$ & $\begin{array}{c}
1+1.25i\\
1.05+0.9i
\end{array}$ & 
$\begin{array}{c}
2+2.5i\\
2.1+1.8i
\end{array}$ & 

$\begin{array}{c}
4+5i\\
4.2+3.6i
\end{array}$ \\

\hline
MSE & 
$2.78\times10^{-4}$ & 
$9.96\times10^{-7}$ & 
$2.99\times10^{-8}$ \\

\bottomrule
\end{tabular}
\label{tab_mse}
\end{table}

\subsection{Performance Comparisons}
\begin{figure}[!t]
\centering
\includegraphics[width=2.6in]{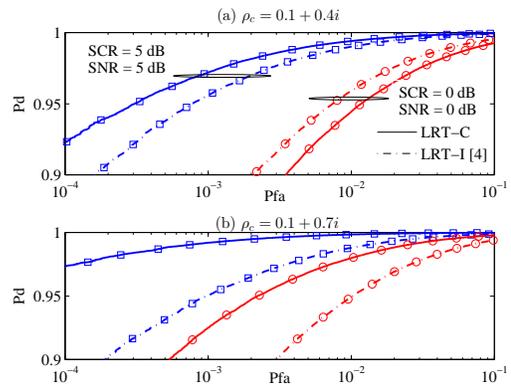}
\caption{Receiver operating characteristic for the proposed LRT-C~(\ref{eq_lr}) and LRT-I in~\cite{dono_2012} with relatively strong target signal ($\mathrm{SCR}=5$~dB, $\mathrm{SNR}=5$~dB) with sample size $Q=5$, denoted by $'\Box'$; and relatively weak target signal ($\mathrm{SCR}=0$~dB, $\mathrm{SNR}=0$~dB) with sample size $Q=20$, denoted by $'\bigcirc'$. (a) $\rho_c=0.1+0.4i$ and (b) $\rho_c=0.1+0.7i$.} \label{fig_roc}
\end{figure}

Here we consider a scenario where the point target has a constant response $T=e^{i\pi/4}$ and the multipath channels are of equal PSD $P_c(\omega_q)=P_c$ over the frequency bands $\{\omega_q\}$. The signal-to-clutter ratio and signal-to-noise ratio are defined as~$\mathrm{SCR}=10\log_{10}\left(|T|^2/P_c\right)$ and~$\mathrm{SNR}=10\log_{10}\left(E_s|T|^2/(Q\sigma_v^2)\right)$, respectively where $E_s=1$. The performance of the proposed LRT detector~(\ref{eq_lr}) under channel correlation, denoted as LRT-C, are evaluated with correlation coefficients $\rho_c=0.1+0.4i$ and $0.1+0.7i$. In addition, we consider the cases where the target has a relatively strong channel response ($\mathrm{SCR}=5$~dB, $\mathrm{SNR}=5$~dB) with sample size~$Q=5$, denoted by the square markers in Fig.~\ref{fig_roc} and relatively weak target response ($\mathrm{SCR}=0$~dB, $\mathrm{SNR}=0$~dB) with $Q=20$, denoted by the circle markers.

Fig.~\ref{fig_roc} shows results from Monte Carlo simulations for the Receiver Operating Characteristic~(ROC), where the detection probability~(Pd) is plotted as a function of the false alarm probability~(Pfa). The ROC of the proposed LRT-C detector is evaluated by test~(\ref{eq_lr}). For comparisons, we also computed the ROCs of the LRT detector designed for independent TR channels (LRT-I)~\cite{dono_2012}. From Fig.~\ref{fig_roc} we can observe that the proposed \mbox{LRT-C} detector outperforms the existing \mbox{LRT-I} by a substantial margin when the target is relatively strong. These results are expectable since the \mbox{LRT-C} detector utilizes the channel correlation, which increases the extent of coherence between the TR signal and the multipath channel. As $\rho_c$ increases to $0.1+0.7i$, the blind TR detection tends to a coherent detection and the proposed LRT-C detector achieves improved detection probability at a fixed false alarm probability by capturing the channel correlation. When the target response is weak with small channel correlation, the performance of LRT-C becomes worse than the LRT-I. This observation is consistent with the Neyman-Pearson theorem~\cite{kay_1993}, as the increased approximation error of~(\ref{eq_f1}) under this condition leads to a non-trivial deviation from this optimal detector.

\section{Conclusion}\label{sec_con}
We proposed a blind time reversal detector which works in the presence of correlated channels. Using Edgeworth expansion, a simple and accurate closed-form approximation was derived for the likelihood ratio test. Simulations show the superiority of the proposed LRT-C detector in scenarios with arbitrary channel correlation and relatively strong target response signal.


%
%

\ifCLASSOPTIONcaptionsoff
  \newpage
\fi



%

\end{document}